%% file: hijacking_repeater.tex
\documentclass[12pt]{iopart}

\usepackage{iopams}  
\usepackage{graphicx}
\usepackage{multirow}
\usepackage{color}
\usepackage{ulem}
\begin{document}

\title{The Network Impact of Hijacking a Quantum Repeater}

\author{Takahiko Satoh, Shota Nagayama, Takafumi Oka and Rodney Van Meter}
\address{Keio University}
\ead{\{satoh, kurosagi, takafumi, rdv\}@sfc.wide.ad.jp}
\vspace{10pt}
\begin{indented}
\item[]\today 
\end{indented}

\begin{abstract}
In quantum networking, repeater hijacking menaces the security and
utility of quantum applications.
To deal with this problem, it is important to take a measure of the impact
of quantum repeater hijacking.
First, we quantify the work of each quantum repeater with
regards to each quantum communication.
Based on this, we show the costs for repeater hijacking
detection using distributed quantum state tomography and the amount of
work loss and rerouting penalties caused by hijacking.
This quantitive evaluation covers both   purification-entanglement swapping
 and quantum error correction repeater networks.
Naive implementation of the checks necessary for correct network
operation can be subverted by a single hijacker to bring down an
entire network. Fortunately, the simple fix of randomly assigned
testing can prevent such an attack.

\end{abstract}
\vspace{2pc}

\input{Introduction}

\section{Quantum repeater networks}
\label{Background}
\begin{figure}[htb] 
\centerline{\includegraphics[width=0.65\textwidth]{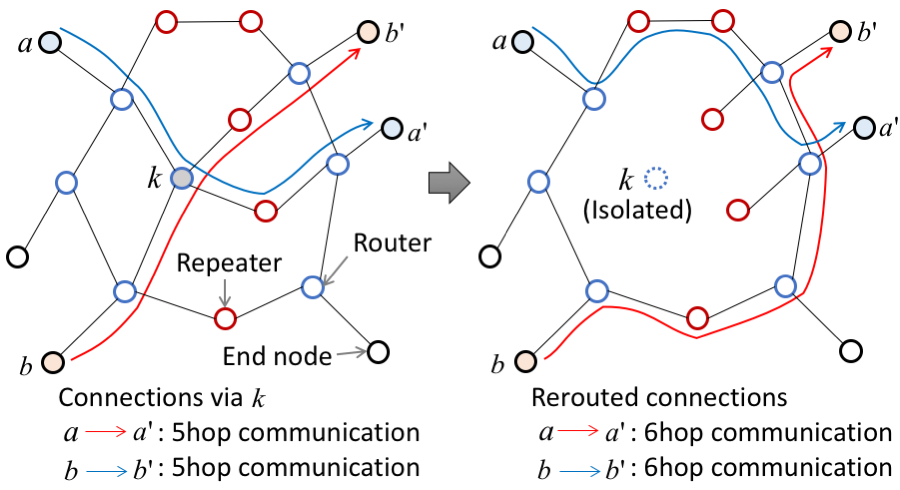}}
\caption{The conceptual view of a quantum repeater network. In this
  diagram, links denote optical fiber and nodes denote quantum repeaters
  serving as Routers, Repeaters, and End Nodes. After detection of the
  hijacking of Router $k$, the  network isolates $k$
  and the connections passing through $k$ are rerouted.} \label{fig:QI}
\end{figure}

Before discussing the effect of repeater hijacking, we show the
basic organization of a quantum repeater network with several active connections
and the isolation of a hijacking repeater in Fig.~\ref{fig:QI}.
Each node corresponds to a quantum repeater and each link corresponds to
a channel such as optical fiber.
When the network detects repeater hijacking, the corresponding repeater is
isolated from the network and connections adopt newly recalculated shortest paths~\cite{van-meter:qDijkstra}.
 
We show the definition of variables for this paper in Table~\ref{tab:var}.
\begin{table}[h]
\caption{\label{tab:var}The correspondence table of variables.}
\begin{tabular*}{\textwidth}{@{}l*{15}{@{\extracolsep{0pt plus12pt}}l}}
\br
$W$ &\multicolumn{1}{p{0.8 \linewidth}}{ Work of the
  entire network. Work denotes total number of attempts to share
  Bell pairs for teleportation per second ~\cite{van-meter:qDijkstra}.}\\
$W_k$, $W'(W'')$ & Work of repeater k,  work including rerouting penalty.\\
$H$ & Amount  of work to create end-to-end Bell pair.\\
$D$ & Data (packet) rate of communication, in Bell pairs per second.\\
$h$ & \# hops between two nodes in a communication session.\\
$L$ & Amount of work loss.\\
$F,~F',~F''$& Initial, purified, twice purified fidelity.\\
$E(F)$ & \multicolumn{1}{p{0.8 \linewidth}}{Expected number of fidelity $F$ base Bell
  pairs consumed to build one final Bell pair.}\\  
$P_{1(2)}$& Success probability of round 1~(2) of purification.\\
$M$& Maintenance state analysis cost, in Bell pairs.\\
$C$& Total network capacity, in Bell pairs per second. \\
$R$ & Maintenance rate, in Bell pairs per second.\\
$S,~S',~S''$ & The slack of network at each time point.\\
\br
\end{tabular*}
\end{table}

\subsection{Quantum repeater}
A quantum repeater is a device for quantum communication capable
of local quantum operations and conservation of quantum information.
Quantum repeaters perform quantum communications using shared Bell
pairs between target repeaters.
The implementation model of a repeater is classified according to the Bell
pair sharing method, as discussed in Sec.~\ref{Introduction}.
A repeater's four main responsibilities are: 1. link-level entanglement
creation; 2. connection of quantum states for multi-hop communication;
3. management of errors; and 4. participating in the management of the network.

In this paper, we distinguish the types of quantum repeaters based on
their connectivity.
A {\it Router} is connected to three or more links.
A {\it Repeater} has exactly two external links, so that it is useful in
a line only.
An {\it End node} has one link and is connected to a network constructed of 
Routers and Repeaters.
Our investigation focuses on Repeater and Router hijacking,  because
hijacking of an End node allows the hijacker to report any results to
the application, up to and including completely forging the existence
of the quantum network itself.~(See Sec.~\ref{capabilities}). End nodes also cannot
spread malicious effects to surrounding networks.

In a quantum repeater network,  connections are multiplexed~\cite{repeater-muxing}.
Each repeater may be  used as the start, end and relay point of multiple
quantum communication sessions simultaneously.
\subsection{Quantum state analysis}
\begin{figure}[htb] 
\centerline{\includegraphics[width=0.65\textwidth]{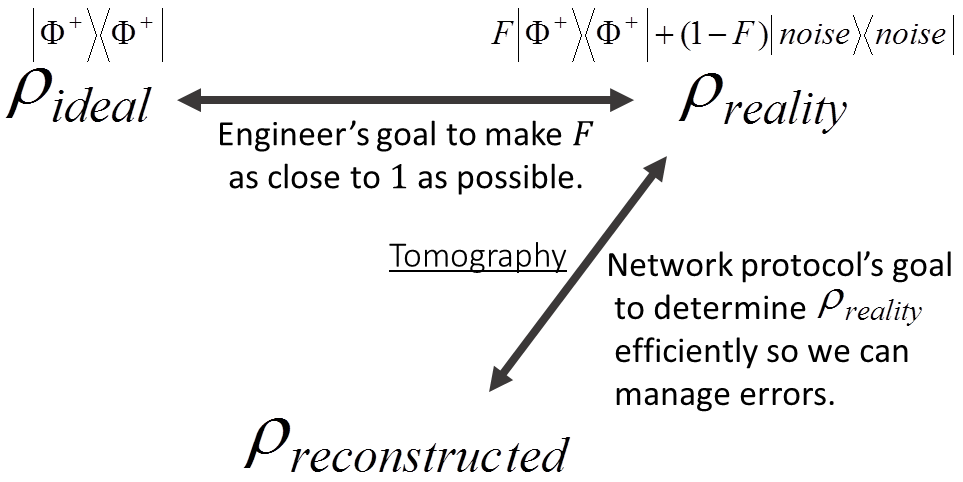}}
\caption{The process of tomography. This figure is shared from~\cite{oka16:qcit}.} \label{fig:statetomo}
\end{figure}

Quantum tomography is a conventional method for estimating the actual quantum
states $\rho_{reality}$ generated by a physical apparatus.
For the purposes of this paper, any state analysis technique will do, but for
concreteness we present our analysis in terms of tomography.
As shown in Fig.~\ref{fig:statetomo}, the density matrix
$\rho_{ideal}$ cannot exist in reality, since errors occur in
any physical apparatus. Utilizing quantum tomography, a large number of measurements of
$\rho_{reality}$ give us the fidelity between $\rho_{ideal}$ and $\rho_{reconstructed}$.
 This scheme can be applied to confirming the operational fidelity of an optical 
fiber and repeater setup, end-to-end, as well as to the detection of repeater hijacking.

\subsubsection{Quantum key distribution}
QKD has the ability to detect eavesdroppers~\cite{bb84}.
For a noisy channel, information reconciliation to detect errors and privacy amplification to reduce the amount of information divulged by information reconciliation are proposed, at the cost of reducing the generated secret key size~\cite{Brassard94secret-keyreconciliation,doi:10.1137/0217014}.
Acting as a simplified form of state analysis, QKD can find evidence of 
eavesdroppers operating on the quantum states. 
However, the simplest form assumes that the hijacker is measuring or completely
entangling with every quantum state. If the hijacker is more
sophisticated and attacks fewer states or entangles only weakly, the effects
are more subtle and require more careful checks to detect.

If our only goal is to prove a Bell inequality violation, the CHSH
  game is good enough and requires fewer quantum resources than
  tomography~\cite{gill2014statistics,bierhorst2015robust,elkouss2016nearly}.
Of course, quantum tomography can more thoroughy evaluate the quantum channel state
 while proving a Bell inequality violation. 
From the point of view of network management, evaluation of the state
created by the channel is an important, fundamental task, to which
quantum tomography is well suited. Rather than add a second protocol
for the narrow purpose of looking for hijacked repeaters, it makes
sense to combine the two tasks, so we propose the adoption of quantum
tomography for hijack analysis as well.
 
\subsubsection{Distributed style tomography}
To use tomography on a repeater network, we have suggested a distributed
quantum state tomography protocol using TCP/IP networks to exchange the
classical information about measurement bases and results, and the
output of the tomographic calculations~\cite{oka16:qcit}. 
In that protocol, we analyze tomography between remotely located
repeaters using Bell pairs that have been purified twice.
Against Werner states $\rho$ with fidelity $F$,
\begin{eqnarray}
\rho = F \vert \Phi^{+} \rangle \langle \Phi^{+} \vert 
+ \frac{1-F}{3}\left(\vert \Phi^{-} \rangle \langle \Phi^{-} \vert 
+\vert \Psi^{+} \rangle \langle \Psi^{+} \vert 
+\vert \Psi^{-} \rangle \langle \Phi^{-} \vert \right),
\end{eqnarray}
as input states, the second round of
  purification~\cite{bennett1996purification,deutsch1996quantum} gives us a greater boost in fidelity than the first
  round~\cite{van-meter14}.  The use of
  many rounds of purification increases the number of Bell pairs
  consumed, raising the required number of entanglement attempts exponentially.
  To keep the analysis straightforward, we assume without loss of generality
  the use of two rounds.
We give a brief summary of our protocol in Table~\ref{tab:tomo}.
\begin{table}
\caption{\label{tab:tomo}The procedure for checking for the presence
  of a hijacked repeater.}
\begin{tabular*}{\textwidth}{@{}l*{15}{@{\extracolsep{0pt plus12pt}}l}}
\br
\# Steps & \hspace{2mm}Master repeater (Node 1) & Slave repeater (Node 2)\\
\mr
Prep 1.& \multicolumn{2}{c}{Sharing initial Bell pairs with fidelity $F$.}\\
Prep 2.& \multicolumn{2}{c}{Performing two round purification to create final Bell pairs.}\\
\mr
1. & \multicolumn{2}{p{0.7 \linewidth}}{Select Bell pairs for tomography using
  synchronized, secure pseudo random number generators at each end.}.\\
2. & \multicolumn{2}{p{0.7 \linewidth}}{ Choose measurement basis using local pure
  random selector, receive pulse and measure qubit.}\\
3. & \multicolumn{1}{l|}{\hspace{0mm}M:~Receive basis and result from Node 2.} &
S:~Send basis and result to Node 1.\\
Branch: & \multicolumn{1}{l|}{\hspace{0mm}Pulse received on both sides?} & (S:wait)\\
\hspace{8mm}Yes: &  \multicolumn{1}{l|}{\hspace{8mm}$\rightarrow$Recalculate d.m. and Fidelity.} &\\
\hspace{8mm}No: &  \multicolumn{1}{l|}{\hspace{8mm}$\rightarrow$Back to Step 1.}&\\
Branch: &  \multicolumn{1}{l|}{\hspace{0mm}Is fidelity over the threshold? }&\\
\hspace{8mm}Yes: &  \multicolumn{1}{l|}{\hspace{8mm}$\rightarrow$Send d.m. to Node 2. }& S:~Receive d.m. from Node 1.\\
\hspace{8mm}No: &  \multicolumn{1}{l|}{\hspace{8mm}$\rightarrow$Back to Step 1.}&\\
Finish. &\multicolumn{2}{p{0.7 \linewidth}}{ Terminate tomography, move link to production use.}\\
\br
\end{tabular*}
\end{table}
Like QKD, not only must the choice of Bell basis for measurement be
random, but \emph{which Bell pairs} in the sequence of generated
states are allocated to tomography must appear random; if e.g. every
tenth Bell pair is used for hijacker detection, she will simply avoid
hijacking those states.
Doing the selection purely randomly and
independently at
each node will be very inefficient; the two nodes will choose the same
pairs with only probability $P^2$. Instead, operation of this
detection process will proceed using pseudorandom selection of states,
which requires properly synchronized, cryptographically secure
pseudorandom number generators between each pair of repeaters
performing hijack monitoring.
The setup of this generation process is
beyond the scope of this paper. 

\begin{figure}[htb] 
\centerline{\includegraphics[width=0.65\textwidth]{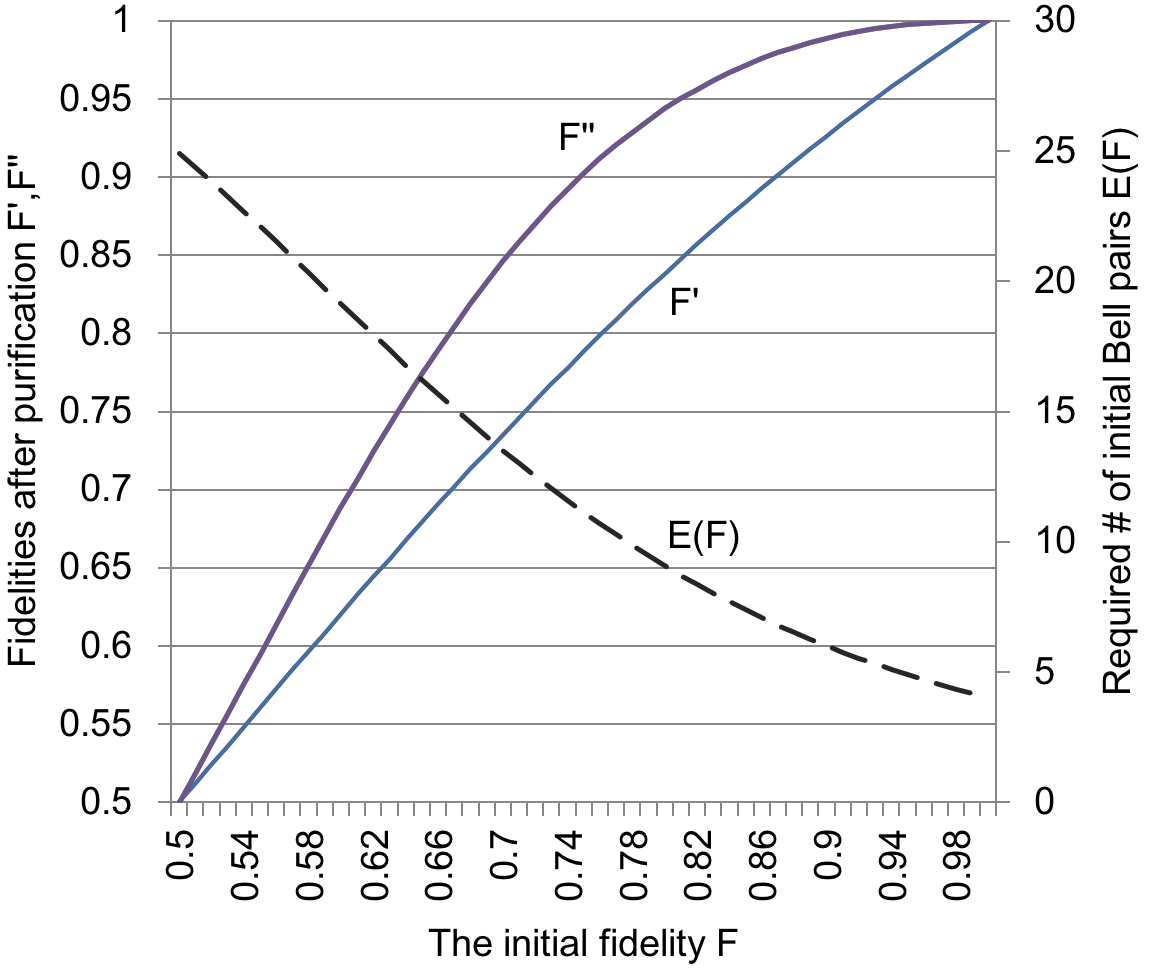}}
\caption{The expected of $\#$ of Bell pairs $E(F)$ and purified
  fidelities for one round ($F'$) and two rounds ($F''$) of
  purification, as a function of the initial fidelity
  $F$. Here, local operation and measurement are considered perfect.} \label{fig:purifi}
\end{figure}

\begin{figure}[htb] 
\centerline{\includegraphics[width=0.65\textwidth]{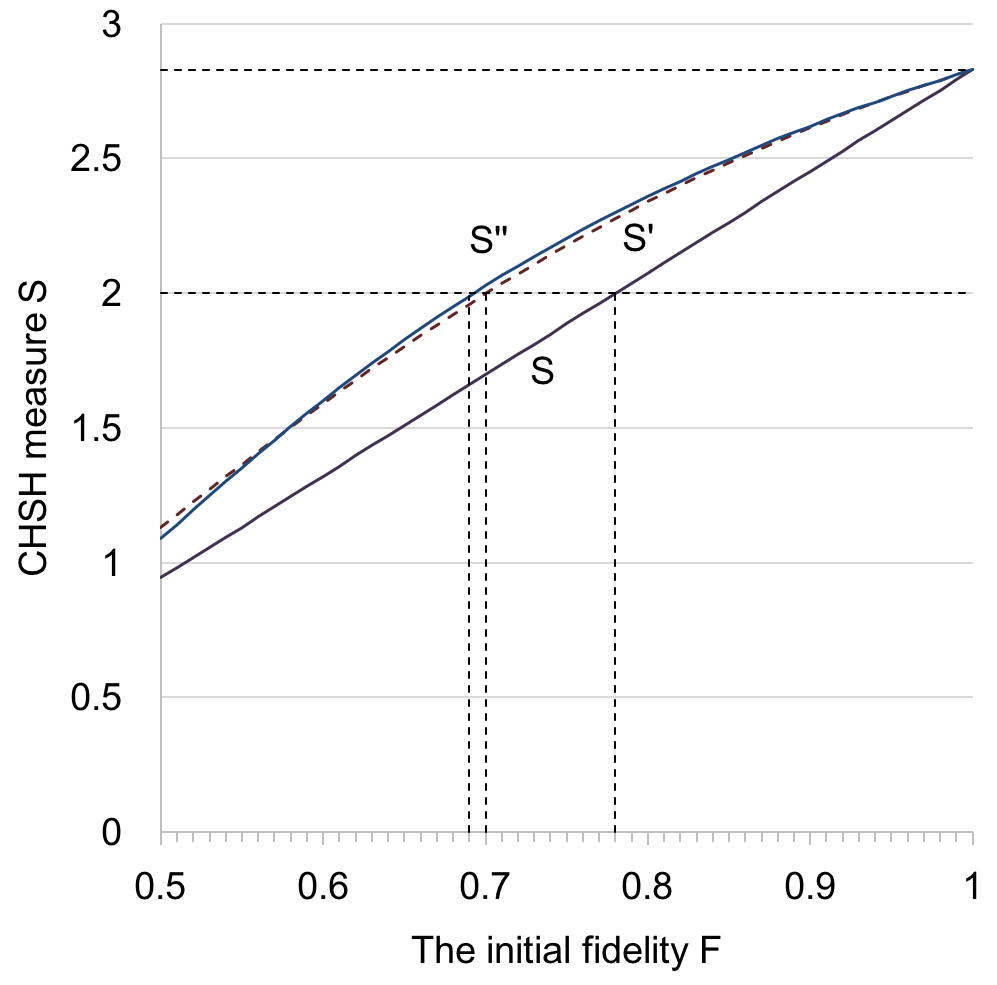}}
\caption{The CHSH measure $S$ as a function of the initial fidelity
  $F$. The lower curve denotes the measure of initial Bell pairs. The
  upper~(dashed) curve denotes the measure of twice~(once)-purified
  Bell pair $S''~(S')$. The upper and lower horizontal dashed line
  denote Tsirelson's bound~($S=2\sqrt{2}$) and the lower bound for
  violations of Bell's inequality~($S=2$). Purified Bell pairs violate
  the bound at about 
$F=0.7$. Non-purified Bell pairs violate the bound at about $F=0.78$. } \label{fig:CHSH}
\end{figure}

The ability of tomography to distinguish the presence of malicious
activity from noise depends on the state of the Bell pairs can that be
generated. Operational networks are expected to make extensive use of
purification. We expect that purification will be conducted in pairs
of rounds, first suppressing $X$ errors, then $Z$ errors. Fig.~\ref{fig:purifi} shows
the fidelity after one round ($F'$) and two rounds ($F''$) of
purification. Beginning with Werner states, the first round of
purification shuffles $X$ errors into $Z$ error terms, with minimal
improvement to fidelity.  The second round of purification then
produces a large improvement in fidelity. 
Detection of malicious activity depends on our ability to demonstrate
violation of the CHSH inequality.  Fig.~\ref{fig:CHSH} shows the expected CHSH
measure with the initial Werner states ($S$), after one round of
purification ($S'$), and after two rounds ($S''$).
Interestingly,
although the fidelity increase after one round of purification is
small, the asymmetric error terms produce a much larger gap in the $S$
value.  Therefore, two rounds of purification are desirable for Bell
pairs destined for application use, but one round is preferred for
Bell inequality violation.  For further details on the behavior of
purification and CHSH, see~\ref{app_puri}.

\subsection{Network administrator}
\label{network_administrator}
To discuss the identification of hijacking repeater, we assume the
existence of a central
network administrator who has his own trustworthy nodes and collects
all end to end tomography results.
When hijacked node is specified, the administrator promptly isolates the
involved node and delivers updated network routing tables that do not use
the isolated node.

\subsection{Phases of network operation}
\label{Phases_no}
In this research, we classify the phases of network operation as follows.
\subsubsection*{Phase 1. Network bootstrapping.}
At the start of network operations, we need to initialize network components.
To check the condition of quantum repeaters and links, some types of
tomography (which are explained in Sec.~\ref{tomographytype}) are utilized.  
Almost the entire capacity of the network is spent to execute these operations, so
that quantum communications for users are not yet provided.
Based on the information acquired via this tomography, the network nodes create routing
tables for selecting paths through the network.
\subsubsection*{Phase 2. Normal operation.}
In normal operations, the network performs quantum communications for
end node applications and
various tomography operations for the maintenance of the network.
The network slack prevents instability of
connections and requires us to minimize possible maintenance costs.
This phase is the main portion of network operations and continues until the
detection of repeater hijacking.
Dynamic changes to traffic and topology occur, but for our purposes
here are  treated as static.
\subsubsection*{Phase 3. After repeater hijacking detection.}
The amount of useful work lost depends on our lag in detecting the
start of hijacking, which in turn depends on the frequency of
tomography.
The network performs rerouting operations and isolates all suspected repeaters.

Reduced network performance and increased communication costs shrink
the slack.

\subsubsection*{Phase 4. Return of innocent repeaters}
By careful verification, if the administrator identifies the actual
hijacked repeater, he returns all isolated innocent repeaters to
the network.
After these operations, the network is reset to a new steady-state
equilibrium, giving us a new Phase 2.

\section{The impact of repeater hijacking}
\label{Impact}

\subsection{The Hijacker's Capabilities}
\label{capabilities}
For our model of a hijacker, we assume that 
the hijacker has taken complete classical control of a quantum
repeater, but is not capable of new physical actions such as blinding
detectors~\cite{gerhardt2011full}.  
We are focusing on the hijacker's selective use of the repeater's
quantum capabilities, such as modifying the circuits used for
purification, quantum error correction or entanglement swapping.
The hijacker may alter destinations of operations, rerouting
connections or entangling additional qubits to quantum states that are
being developed. Those qubits may be local within the hijacked
repeater, or may involve a fourth party elsewhere in the network.
Accordingly, the hijacker is able to foil quantum communications and
potentially to steal quantum information.

Hijacking of an End node would give the hijacker the
  ability to convince an application of any behavior by the network,
  just as taking control of a personal computer's operating system
  allows a hacker to emulate any behavior by the system. This means
  the hijacker can conceal hijacking all the way to the end of the application's
  work. Such an all-powerful hijacker is beyond the scope of this work.

 In this paper, we focus on the detectable hijacking of Routers
and Repeaters.
The hijacker's goal is to maximize the hijacking time and the range of
influence while remaining undetected. 

In following parts of this section, we define the work of a network,
classify two types of effects of repeater hijacking, and quantify the
work loss problem.

\subsection{Defining  connection cost}
To quantify the network impact of the hijacking of a quantum router,
we need to understand the amount of work~(see the
  work entry in Table~\ref{tab:var}) done by the network.  We can
either count connections and sum the work done per connection across
the entire network, or we can count nodes and sum the work done per
node across the entire network.  As our goal is to study the hijacking
of a single router, the latter may seem more desirable, but the
behavior of ES connections makes the former easier in some ways.

First, let us establish a definition of \emph{work}.  The primary
element of work in a quantum network is the quantum optical pulses
used to create entanglement across a link.  The work done for a
connection, then, is the total number of pulses used on behalf of
that connection across the set of nodes
 involved~\cite{van-meter:qDijkstra}.

For one connection achieving an end-to-end Bell pair generation rate
(data rate) of $D$ Bell pairs/second consumed by the application
such as QKD or quantum distributed computation, how
much work is done?  The work done at each node is $D$ multiplied by an
overhead factor $H$ that depends on the type of repeater network:
\begin{eqnarray}
\mbox{ES type repeater:~}&H^{ES}_{k,i} &= O((h_{k,i})^c),\\
\mbox{QEC type repeater:~}&H^{QEC}_{k,i} &= O(d),
\end{eqnarray}
where $h_{k,i}$ denotes the path length for connection $i$ passing
through the repeater $k$,
 and $c$ is a small constant dependent on the details of the purification scheme.
$d$ denotes the code distance or the block size for error correction.
(For comparison, because the error correction overhead is essentially
 constant, the amount 
of work done at a node in a classical network is simply linear in the
number of bits being transmitted, $H^{C}_{k,i} = O(1)$.)

To calculate the work at repeater $k$, we simply sum over the $HD$
product for each connection passing through the repeater,
\begin{eqnarray}
W_k = \sum_{i\in connections} H^{type}_{k,i}D_{k,i}.
\end{eqnarray}
Note that this sum is only the work contributing directly to
end-to-end states to be consumed by application, and does not include
the cost of tomography.
Then, we can quantify the work of network $W$ as follows: 
\begin{eqnarray}
\label{def:work}
W = \sum_{k\in nodes} W_{k} = \sum_{k} \sum_{i\in connections}
H^{type}_{k,i}D_{k,i}. 
\end{eqnarray}

\subsection{Quantifying work loss}
When we detect the hijacking of repeater $k$ with duration $t$, we can
consider all connections using $k$ to be lost. The amount of work loss $L_k$
is more than $W_k$, with other repeaters work corresponds to lost
connections as follows:
\begin{eqnarray}
L_k t=  \sum_{i\in connections} (h_{k,i} +1)H^{type}_{k,i}D_{k,i} t.
\end{eqnarray}

\section{Preventing  repeater hijacking}
\label{Preventing}
\subsection{Distributed style tomography for hijacking detection}
\label{MITM}
Tomography is a conventional scheme for characterizing a quantum
state~\cite{altepeter2005photonic}, and can be repurposed to detect quantum state falsification. 
This scheme can be applied to verifying the link state and the repeater
state, including the detection of repeater hijacking.
For example, the hijacker entangles a third qubit $C$ with the Bell
pair $\vert \Phi^{+}\rangle_{AB}$ Alice and Bob are sharing.
The system $\rho^{ABC}$ without $C$ becomes a completely mixed state as follows:
\begin{eqnarray}
 \rho^{ABC} &=& (\vert 000 \rangle + \vert 111
 \rangle)\otimes(\langle 000 \vert + \langle 111 \vert),\\
\rho^{AB} &\equiv&{\rm Tr}_{C}(\rho^{AB})\\
&= &  \vert 00  \rangle  \langle 00\vert_{AB} + \vert 11
\rangle \langle 11 \vert_{AB}.
\end{eqnarray}
Then, an ensemble of $\rho^{AB}$ cannot prove a Bell inequality violation and
hijacking can be  detected statistically over a set of trials.
Of course, this works only if both sides share measurement results
securely.
If the eavesdropper can modify or control both the quantum and
classical connections between the two parties, she can send false
measurement results and fake a Bell inquality
violation~\cite{ekert1991qcb,terhal2004entanglement,ekert2014}.
To avoid such a man in the middle attack, we assume all classical
communications are authenticated and unmodified. Then, each node can
send classical messages to other nodes securely.

\subsection{Work for tomography}
\label{tomographytype}
Next, we discuss the work for a node and any type of link tomography.

\subsubsection*{Node tomography (self check)}
State tomography and process tomography in each node are requisite for verification of gate
operation and internal self functions. This tomography cannot detect
repeater hijacking.
In this research, we assume this operation is done periodically.

\subsubsection*{$1$~hop link tomography}
Periodic but frequent tomography between nearest neighbor
repeaters is requisite for verification of the state of each link.
In prior work~\cite{oka16:qcit}, we have shown that we need around
$2000\sim 3000$ Bell pairs for
tomography to reconstruct the state with $99$\% fidelity for initial Bell pair fidelity of
$F\simeq 0.65 $.
We can define the tomography work for link $j$ as follows:
\begin{eqnarray}
 M^{link}_{j} = B(F)E(F),
\end{eqnarray}
where $B(F)$ denotes required number of Bell pair for tomography based
on initial Bell pair fidelity $F$ and  $E(F)$ is as in Eq.~(\ref{expbp}).

\subsubsection*{Multilevel recursive tomography for ES model}
In ES model repeater networks, multilevel recursive tomography is 
requisite for checking the condition of purification, entanglement
swapping operations and the connection state.
The tomography work at ES model repeater $k$ for connection $i$ becomes as follows:
\begin{eqnarray}
M^{con(ES)}_{k,i} = \underbrace{B(F)E(F)\!B(F'')E(F'')\!B(F'''')E(F'''')\!\cdots
  \!B(F^{'\cdots '})E(F^{'\cdots '})}_{log({h_{k,i}})}.
\end{eqnarray}
We can calculate $F^{'\cdots '}$ using
Eq.~(\ref{ffid}) recursively.

\subsubsection*{End to End tomography for QEC model}
During quantum communications on QEC model repeaters, both end
nodes need to execute tomography periodically to check the
connection state. 
If done with appropriate attention to security,
this operation can detect the repeater hijacking, but cannot identify
which repeater has been hijacked if the connection length is larger
than $2$ hops.
The tomography work at QEC model repeater for connection $i$ becomes as follows:
\begin{eqnarray}
M^{con(QEC)}_{k,i} &=&  h_{k,i}B(F) E(F).
\end{eqnarray}

We will discuss the frequency and the total work of
those tomography in following section.

\subsection{Identification of hijacking repeater}
When the hijacked repeater always acts maliciously on every
connection, the administrator can identify that repeater by using the
combination of reported tomography results.
In contrast, when the hijacked repeater targets only the connection between
two specific repeaters, whether administrator can identify or not depends on the
repeater models.

For example, in Fig.~\ref{fig:QI}, if it is known that the $a-a'$ and
$b-b'$ connections both pass through a hijacked repeater, we can infer
that the malicious node is $k$.
Naturally, this identification process requires substantial support
from the classical network protocols.
\subsubsection{ES model}
\begin{figure}[htb] 
\centerline{\includegraphics[width=0.65\textwidth]{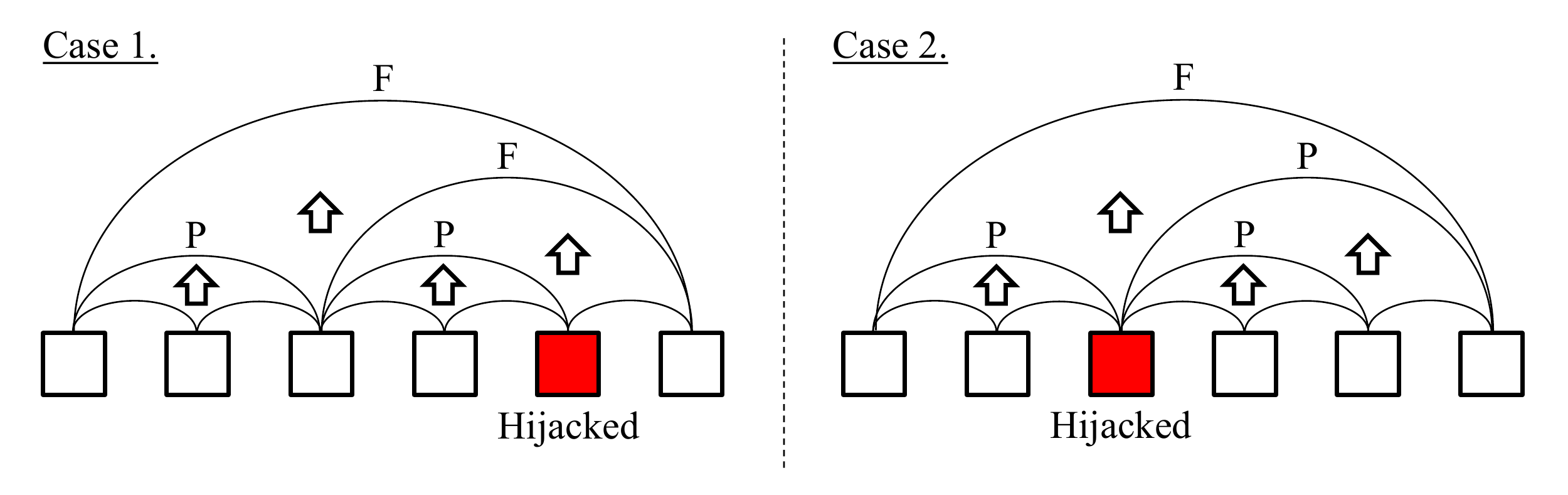}}
\caption{Arrows denote entanglement
  swapping operation by the repeater below. P and F denote tomography
  results whether pass or fail. By process of elimination, we can
  identify the hijacked repeater.} \label{fig:idenes}
\end{figure}

In the ES model, repeaters perform entanglement swapping in a nested tree
to share final Bell pairs for communication.
As shown in Fig.~\ref{fig:idenes}, if we perform cryptographically
secure  tomography along with  every entanglement swapping operation,
we can finger the culprit repeater.
After the identification, the network should isolate the hijacked repeater
as soon as possible and reroute connections passing through that repeater. 
These changes  reduce total network performance and increase
communication costs so that the slack of network is suppressed.
In exchange for this burden, the effects of the hijacking are rooted out.

\subsubsection{QEC model}
\label{qec_model}
\begin{figure}[htb] 
\centerline{\includegraphics[width=0.65\textwidth]{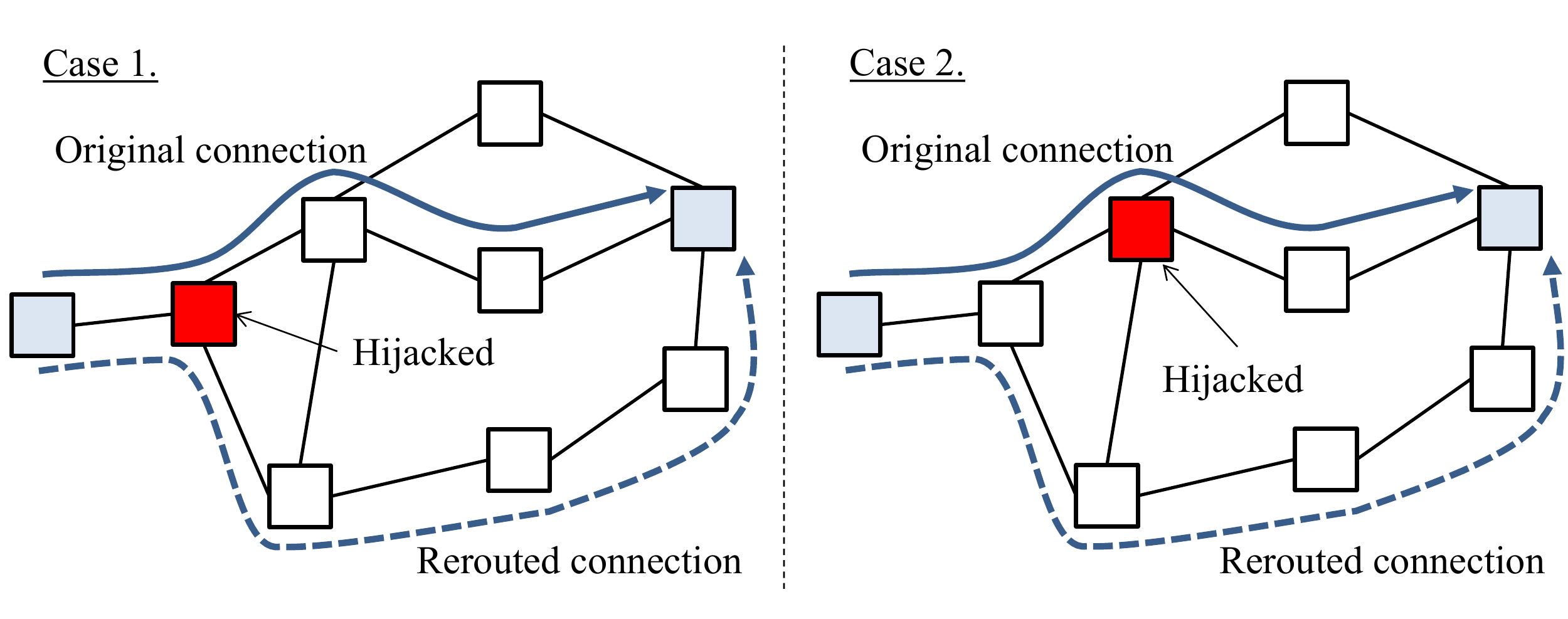}}
\caption{The information from hijacking detection process narrows down the
  candidates. The rapid sharing of this information across the entire network is
  important. } \label{fig:idenqec}
\end{figure}

In the QEC model, we can detect the repeater hijacking using end to end
tomography but our ability to correctly identify the hijacked repeater
is weak.
When the hijacked repeater targets specific one connection through
multiple repeaters, we cannot identify the hijacked repeater.
Then we must reroute the connection and
should temporarily increase the frequency of tomography to determine
if we have successfully rerouted around the hijacked repeater.
As shown in Fig.~\ref{fig:idenqec}, if the attack continues,
identification of hijacked repeater may be possible.
However, in the example in Case~2. with no ability to route around the
hijacked node, the connection may have to be abandoned.
The aftermath of the identification and effects associated with it are
similar to the  ES model.

\subsection{The appropriate frequency of tomography.}
To maintain the integrity of a repeater network, we need periodic tomography
on each link and node. Here, we discuss the
appropriate frequency of tomography.

\begin{figure}[htb] 
\centerline{\includegraphics[width=0.65\textwidth]{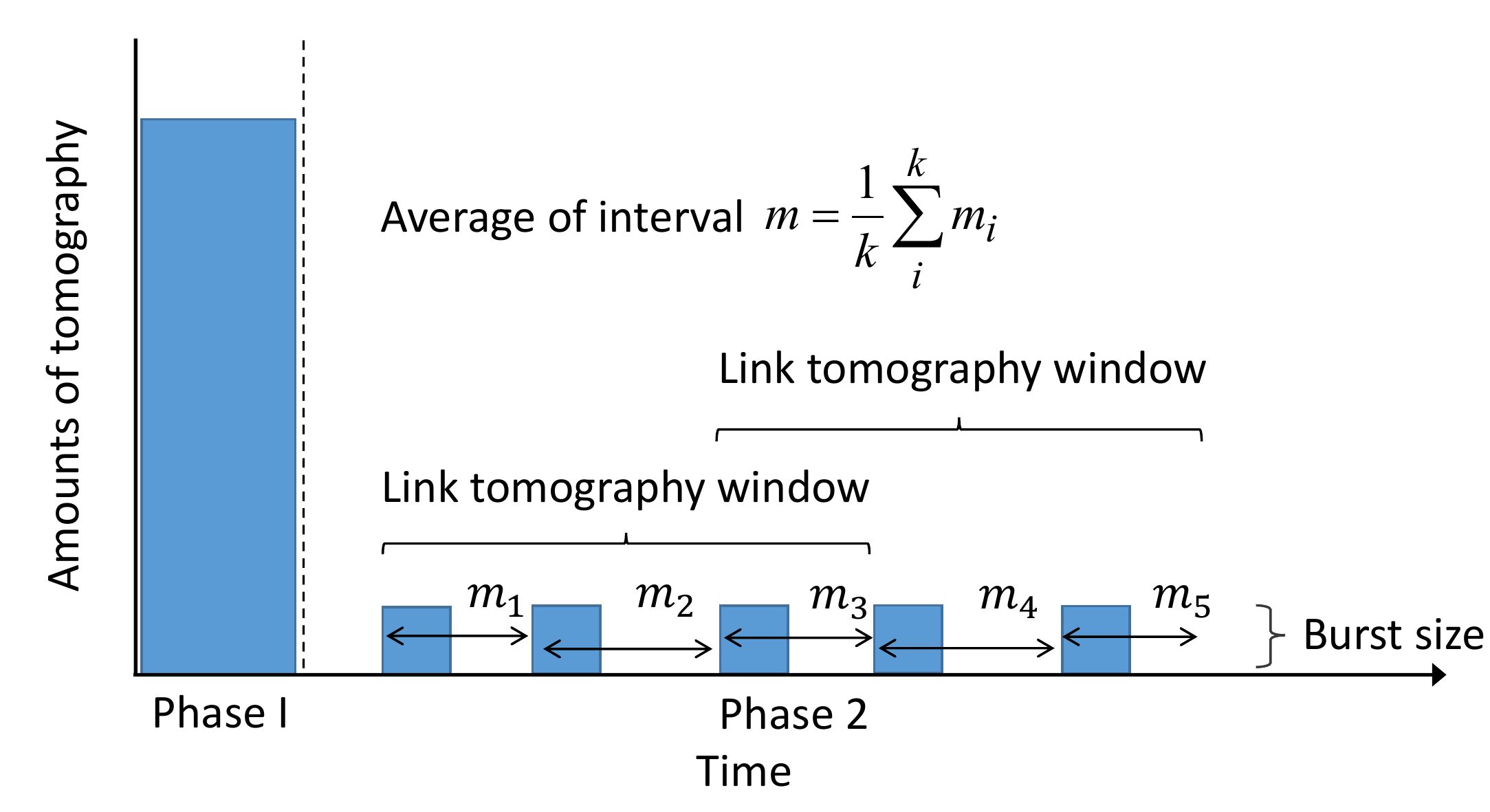}}
\caption{In normal network operation (Phase 2), we perform the
  burst size of tomography to each link with average interval $m$. We can
  verify the state of link using constant number of sequential
  tomography results. In this figure, the window size is $3$.} \label{fig:Tomo_window}
\end{figure}

To provide load balancing, sliding window is a suitable
scheme for link tomography. Each link performs burst size tomography
operations with average interval $m$. Here, each interval must vary
randomly to foil attempts by the hijacker to remain undetected by
laying low and not intercepting states to be used for tomography.
The network can verify the link states
using the tomography results of the prescribed window, as shown in
Fig.~\ref{fig:Tomo_window}. 
There is the following relation between the time for verification $T$ and
the window size for tomography $w$:
\begin{eqnarray}\label{eq:Tomo_window}
T = wm.
\end{eqnarray}
The total tomography cost for network maintenance per second $R$ can
be described as follows: 
\begin{eqnarray}\label{eq:Tomo_cost}
R = \sum_{j\in links}\frac{M^{link}_{j}}{wm} + \sum_{k\in nodes}\frac{M^{con}_{k}}{w'm'},
\end{eqnarray}
where $w~(w')$ and $m~(m')$ denote the window size and interval of link
(connection) tomography.
For example, if each burst is  $10$~Bell pairs with the interval $m=1$
sec and $2000$~Bell pairs are required for tomography, we need $200$ seconds for
link verification.

\section{Framing innocent repeaters}
\label{sec_framing}
\subsection{Framing an innocent repeater}
If the hijacker continues attacking in the most straightforward
fashion, we can identify the hijacked repeater or prevent attacking as
just described.
However, if the hijacker can identify Bell pairs that will be used for hijack
detection in any layer above his position in the swapping hierarchy or
any connection he terminates, he can substantially damage operation of
large swaths of the network by framing innocent repeaters, convincing
network administrators that other repeaters besides itself have been hijacked.

The hijacker can frame another, innocent repeater in one of two ways:
first, when it is the endpoint of a Bell pair, it can directly falsify
measurement results, causing the failure of the entanglement checks
that test for the presence of a hijacker, in which case the last node
to perform entanglement swapping will be blamed.  Second, if it knows
the sequence of tests performed by the other nodes after entanglement
swapping, it can selectively choose which Bell pairs to corrupt.

\begin{figure}[htb] 
\centerline{\includegraphics[width=0.65\textwidth]{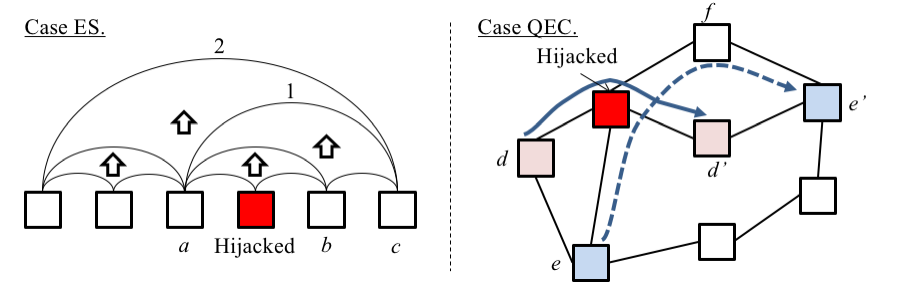}}
\caption{In a path of repeaters in an  ES network, the hijacker can
  frame repeaters $a$ and $b$ in several ways. In QEC a network, the
  hijacker can frame by attacking a specific connection.} \label{fig:Framing} 
\end{figure}

For example, in the left side of Fig.~\ref{fig:Framing}, the hijacked router's
neighbors $a$ and $b$ will check the Bell pairs created after the
hijacker's entanglement swap.  If they detect corruption, they will
naturally blame the hijacker and report him to the network
adminstrator.
If, however, the hijacker allows the $a-b$ Bell pairs
to be created uncorrupted but corrupts the Bell pairs used for the
$a-c$ check (marked 1 in the figure), the $a-b$ check will pass but
the $a-c$ check will fail, and nodes $a$ and $c$ will instead conclude
that $b$ has been hijacked.  The hijacker has successfully framed $b$,
who will now be reported to the network administrator and removed from
the network, pending further investigation.

This level of subterfuge is possible if the hijacker can predict which
Bell pairs will be used for which operations.  For example, if the
first ten Bell pairs created with $a$ are used to check the link, the
next ten to check the $a-b$ connection, and the next ten used to check
the $a-c$ connection, the hijacker knows to leave the first twenty
Bell pairs alone, then corrupt the next ten.

In a QEC network, checks are not done in the same nested fashion, so a
hijacker cannot use this technique to frame other repeaters.  However,
because end-to-end checks are done, he can still disrupt any
individual connection. which then forces rerouting of that connection,
as described in Sec.~\ref{qec_model}. QEC network administrators will use these
end-to-end reports to attempt to identify and isolate the hijacker, as
described in Sec.~\ref{network_administrator}.
In the right half of Fig.~\ref{fig:Framing}, the hijacker is carrying two
connections. If he chooses to corrupt the $e-e'$ connection but not
the $d-d'$ connection, an administrator examining the network will
logically conclude that the bad guy is router $f$, and the hijacker
has successfully framed someone other than himself.

\subsection{Possibility of bringing down the network}
\begin{figure}[htb] 
\centerline{\includegraphics[width=0.65\textwidth]{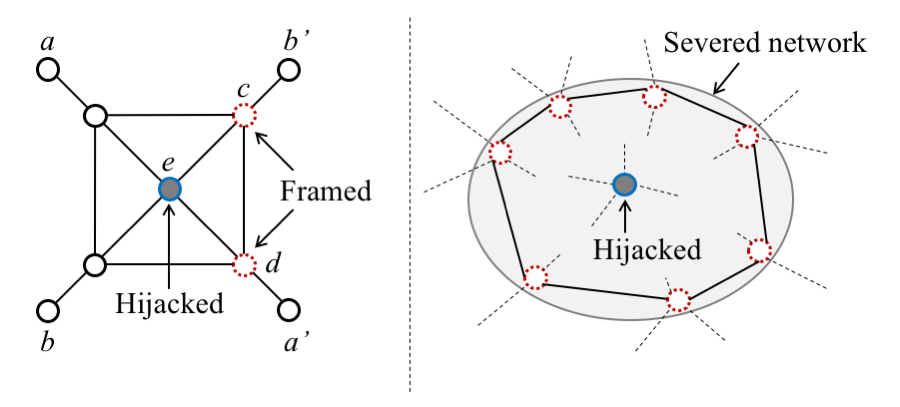}}
\caption{Examples of bringing down the networks by framing. In the left
  case, framing of repeaters $c$ and $d$ by hijacked repeater $e$
  prevents the communications between $a-a'$ and $b-b'$. This example
  shows that the exact framing can bring down a network of a
  particular topology. In the right case, the hijacked repeater severs the network
  surrounded by a circle using framing to several outline repeaters
  (dashed circles) from the entire network. After framing the seven
  repeaters, the network will be partitioned, and nodes inside the
  circle will be unable to communicate with those outside the
  circle.} \label{fig:down} 
\end{figure}

When the attack is detected, the network administrator will isolate
suspect repeaters from the network.
Depending on the structure of the network, framing several carefully chosen repeaters
can bring down the network. We show examples of bringing down the
network  in Fig.~\ref{fig:down}. 
To prevent such a serious situation, the network topology should be
designed robustly.

\subsection{Preventing framing}
Use of a simple or easily predicted sequence of Bell pairs for hijack detection
allows a hijacker to hide his presence and to frame other nodes.
This can result in something of a cat and mouse game if the hijacker
lays low during the investigation.
To prevent this, administrators need to conceal information about the
sequence of hijack detection.
For a given connection the choice of which Bell pairs will be used for
hijack detection can be decided by the end nodes using some form of
secret agreement, or the sequence can be chosen by an administrator
and communicated securely to the set of nodes in the path. The
interval between Bell pairs chosen for hijack detection, at each level
of the check (number of hops spanned by the Bell pairs), must appear
random to any outside observer.
For example, in the left hand side of Fig. 7, node $b$ naturally must
know which Bell pairs shared with $a$ will be used to check for
hijacking between them, but it must \emph{not} be able to predict
which Bell pairs will be used to check for hijacking between $a$ and
$c$.
(both those directly reported by an end node and those that can
otherwise be reasoned to be suspects)

As one of the draft, administrator every give the sequence information to
repeaters which practice just before execution. The execute period
should be to appear in randomly.

Administrator also consider the framing and perform isolation to all
suspect repeaters (reported and all reportable repeaters). 
At the same time as isolation, the administrator should check isolated
repeaters directly to close in on and confirm the identity of the hijacked repeater. After this
quarantine, repeaters shown to be innocent will be gradually returned to
the network. We will discuss the transition of network performance
during these operation in next section.

\section{Post process of hijacking detection}
\label{sec_postprocess}
\subsection{An increase of the work from rerouting penalty}
When the network detects the hijacking of repeater $k$, $k$ and other
suspected repeaters are isolated from other repeaters and all
connections via suspect repeaters are forced to reroute. 
If all connections selected the shortest path, rerouting increases
each path length by $\Delta_{sus,i}$, 
where $i$ denotes connection $i$ passing through suspected repeaters.

The increased work $W'$ in contrast with $W$ in
Eq.~(\ref{def:work}) can be described as follows: 
\begin{eqnarray}
W' =  W - L_{sus} + \sum_{i\in rerouted-connections} (h_{sus,i} +1 +\Delta_{sus,i})
H'^{type}_{sus,i}D'_{sus,i}, 
\end{eqnarray}
where $L_{sus}$, $H'^{type}_{sus,i}$ and $D'_{sus,i}$ are work loss from
isolation of suspected repeaters, the updated work and datarate for
each rerouted connection not passing through suspected repeaters. 

\subsection{The occurrence of Work shedding}
We next investigate the phase of network operations from the point
view of the change of change in the network slack.
\subsubsection*{Phase 1. Network bootstrapping}
During this phase, all network performance is spent on node and link
tomography. 
Normal quantum communication has not started yet.
The slack of the network will be narrow in this phase.

\subsubsection*{Phase 2. Normal operation}
In the normal operation phase, the network requires node and link tomography within
the interval $m$. Quantum communications subject to additional
multilevel recursive tomography cost $M^{con}$ (in ES repeater cases; QEC network requires end to end
tomography) corresponding to communication work.
Then, the network slack $S$ becomes as follows:
\begin{eqnarray}\label{eq:slack}
S = C - W - R.
\end{eqnarray}
Here we assume $W=0.7C$, $R=0.1C$ and $S=0.2C$.
Using the quantitative assumption after Eq.~(\ref{eq:Tomo_cost}), this
network is still provides 20 Bell pairs per second as surplus capacity.

We assume the repeater hijacking event occurs at a certain point in
time in this phase.
For efficient operation, the maintenance cost should be
sufficiently smaller than the worst assumption of the work loss $L^{worst}_{sus}$ as follows:
\begin{eqnarray}
R \ll L_{sus}.
\end{eqnarray}
Here, ``the maximum time from the start to detection of  repeater hijacking''
 is equal to the time for verification $T$.
Using Eq.~(\ref{eq:Tomo_window}), we can consider both the maintenance
state analysis cost $M$ and $L_{sus}$ to be functions of $m$. 
To satisfy this requirement, we should not adopt an excessively small $m$.
A larger $m$ suppresses the maintenance cost but increases the
impact of repeater hijacking.

\subsubsection*{Phase 3. After repeater hijacking  detection}
After repeater hijacking detection, the troubled repeater is purged
from the network and connections through that repeater are rerouted.
As the rerouting progresses, the amount of work lost at a given moment
is gradually reduced and the costs of rerouted connections increase,
as above. 
Then, the slack of the network becomes as follows:
\begin{eqnarray}
S' = (C -C_{sus}) - W' - (R -R_{sus}),
\end{eqnarray}
where the subscript $sus$ denotes all suspected repeaters including the hijacked
repeater $k$ and $R_{sus}$ denotes the maintenance costs for $sus$ and links to $sus$.
As $S$ declines, queueing delays go up. If $S$ becomes zero or
negative, we have insufficient capacity to support our current
work, and must shed some work.
The available network bandwidth goes down and we cannot
avoid communication delay.

As an example, assume the extreme case in which $10$\% repeaters of network are suspected of having been.
Then, we can set $C_{sus}=0.1C$ and $R_{sus}=0.1R$.
Using the assumption after Eq.~(\ref{eq:slack}), in order to keep the
slack value positive, we need to keep $W' <0.81C$.
In other words, the increase in communication cost $W'-W$ accompanying the
route change has to stay less than $0.11C=$ 11 Bell pairs per sec.
As in the above discussion, the slack of the network depends on interval
parameters of tomography and the rerouting penalty.

\subsubsection*{Phase 4. Return of innocent repeaters}
By careful verification, if the administrator believes she has
reliably identified the really hijacked repeater $k$, isolated
innocent repeaters are returned to the network.
Then, the recoverd slack $S''$ and reconfigured work of the network $W''$ becomes as follows:
\begin{eqnarray}
S'' &=& (C -C_{k}) - W'' - (R -R_{k}), \\
W'' &=&  W - L_{k,i} + \sum_{i\in connections} (h_{k,i} +1 +\Delta_{k,i})
H'^{type}_{k,i}D'_{k,i}.
\end{eqnarray}
Since $W''$ is obviously smaller than $W'$, 
in some cases, we can expect that the workload shedding will be solved
by returning the slack to a positive value.

\section{Discussion}
\label{Discussion}
In this paper, we show the effects of repeater hijacking and appropriate network
operations from the viewpoint of the slack on ES and QEC model quantum repeater
networks. 
To make quantitative discussions, we quantify works of repeater
network based on the required number of pulses for distributed quantum tomography.
We present actual phases of quantum repeater network operation and
expected maximum duration of repeater hijacking. 
To suppress work losses, we need larger maintenance costs and network
slack is tight.
For this reason, we recommend that the interval may be adjusted carefully
to prevent the possibility of work shedding and the bringing of huge
work losses for the actual network design.
We also show the difference between ES and QEC model in identification
of hijacking repeater. In the QEC model, we can detect hijacking but
cannot always identify the hijacking repeater. 

A single hijacked node can, in theory, bring down an entire network by
leveraging two  factors to frame other repeaters: the necessity of
many distributed checks in the network to support effective operation,
and the \emph{delayed} dependence of the results of those checks on
prior operations.  Classical networks of course are vulnerable to
various attacks on their routing algorithms that can make portions of
the network unreachable, but no checks equivalent to post-entanglement
swapping tomography are conducted, so no similar subversion mechanism
is possible.  Fortunately, the relatively simple fix of choosing Bell
pairs for hijack checks in a random, secure manner can alleviate this vulnerability.

\section*{Acknowledgements}
We would like to thank Shigeya Suzuki and Yohei Kuga for useful discussion.
This material is based upon work supported by the Air Force Office of
Scientific Research under award number FA2386-16-1-4096.

\section*{References}
\bibliography{satoh_quantum}

\appendix
\section{Purification and CHSH values}
\label{app_puri}
We adopt the following Werner state  $\rho$ as given initial noisy Bell pairs
\begin{eqnarray}
\rho = F \vert \Phi^{+} \rangle \langle \Phi^{+} \vert 
+ \frac{1-F}{3}\left(
 \vert \Psi^{+} \rangle \langle \Psi^{+} \vert 
+\vert \Psi^{-} \rangle \langle \Psi^{-} \vert
+\vert \Phi^{-} \rangle \langle \Phi^{-} \vert 
 \right).
\end{eqnarray}
Using purification protocol between two Bell
pairs~\cite{bennett1996purification,deutsch1996quantum}, 
we first suppress $X$ errors on one Bell pair (which we call subject) while consuming
tool Bell pairs. The correspondence output states,
subject and tool Bell pairs is as follows:
\begin{table}[htb] 
 \begin{center}
  \begin{tabular}{c|cccc}
Source \textbackslash Target & $\Phi^{+}$ & $\Psi^{+}$ & $\Psi^{-}$ & $\Phi^{-}$\\ \hline 
$\Phi^{+}$ & $\Phi^{+}$ & $\times$ & $\times$ & $\Phi^{-}$\\
$\Psi^{+}$ & $\times$ & $\Psi^{+}$ & $\Psi^{-}$ & $\times$\\
$\Psi^{-}$ & $\times$ & $\Psi^{-}$ & $\Psi^{+}$ & $\times$\\
$\Phi^{-}$ & $\Phi^{-}$ & $\times$ & $\times$ & $\Phi^{+}$\\
  \end{tabular} 
 \end{center} 
\end{table}
\\where $\times$ denotes an outcome discarded as failure.
The success probability of this round of purification $P_{1}$ becomes
\begin{eqnarray}
P_{1}&=& \left(F + \frac{1-F}{3}\right)^2 +
\left(\frac{2(1-F)}{3}\right)^2.
\end{eqnarray}
Therefore, the fidelity of once-purified state $F'$ becomes
\begin{eqnarray}
F' &=& \frac{1}{P_{1}}\left(F^2+\left(\frac{1-F}{3}\right)^2\right).
\end{eqnarray}

We next perform purification protocol to suppress $Z$ errors then use two
once-purified Bell pairs as resources. The correspondence of output states,
subject and tool Bell pairs is as follows:
\begin{table}[htb] 
 \begin{center}
  \begin{tabular}{c|cccc}
Source \textbackslash Target & $\Phi^{+}$ & $\Psi^{+}$ & $\Psi^{-}$ & $\Phi^{-}$\\ \hline 
$\Phi^{+}$ & $\Phi^{+}$ & $\Psi^{+}$  & $\times$ & $\times$\\
$\Psi^{+}$ & $\Psi^{+}$ & $\Phi^{+}$  & $\times$ & $\times$\\
$\Psi^{-}$ & $\times$ & $\times$ & $\Phi^{-}$ & $\Psi^{-}$\\
$\Phi^{-}$ & $\times$ & $\times$ & $\Psi^{-}$ & $\Phi^{-}$ \\
  \end{tabular} 
 \end{center} 
\end{table}
\\The success probability of this round $P_{1}$ becomes
\begin{eqnarray}
P_{2}&=& \left(F'+\frac{2(1-F)^2}{9P_{1}}\right)^2 + \left(\frac{2(1-F)(1+2F)}{9P_{1}}\right)^2.
\end{eqnarray}
The fidelity of twice-purified state $F''$ and the expected required number
of initial Bell pairs $E(F)$ becomes
\begin{eqnarray}
\label{ffid}
F'' &=& \frac{1}{P_{2}}\left(F'^2 + \left(\frac{2(1-F)^2}{9P_{1}}\right)^2\right),\\
E(F) &=& \frac{4  }{P_{1}^2 P_{2}}.
\label{expbp}
\end{eqnarray}

To calculate the CHSH measure of Bell pairs, we also adopt the following
usual form of CHSH inequality~\cite{PhysRevLett.23.880}:
\begin{eqnarray}
|S|\leq 2,
\end{eqnarray}
where
\begin{eqnarray}
  S=E(\theta,\phi)+E(\theta,\phi')+E(\theta',\phi)+E(\theta',\phi').
\end{eqnarray}
$\theta$, $\theta'$, $\phi$, and $\phi'$ denote the particular choices
of measurement angle of each qubit. We adopt Bloch sphere angles of $0$, $\frac{\pi}{2}$,
$\frac{\pi}{4}$, and $\frac{3\pi}{4}$ respectively.
\end{document}

%% file: Introduction.tex
\section{Introduction}
\label{Introduction}
Large scale quantum repeater networks will be required for world-wide
quantum communication between arbitrary
nodes~\cite{munro2011designing,van-meter14}. 
The main functionality of the quantum network is to create Bell pairs between  two chosen nodes
~\cite{briegel98:_quant_repeater,6246754}.
Such Bell pairs enable, for example, provably secure shared keys for
classical encrypted communication~\cite{ekert1991qcb},
distributed quantum computation~\cite{
buhrman03:_dist_qc,
buhrman1998quantum,
PhysRevA.89.022317,
broadbent2010measurement,
crepeau:_secur_multi_party_qc,
Broadbent:2009:UBQ:1747597.1748068,
chien15:_ft-blind},
and various kinds of physical experiments~\cite{
PhysRevLett.87.129802,
PhysRevLett.85.2006,
PhysRevLett.109.070503,RevModPhys.79.555}.

Implementations of quantum repeaters are classified into three generations by Muralidharan et al.
~\cite{muralidharan2016optimal}.
(1) The first generation employs entanglement purification
~\cite{briegel98:_quant_repeater,repeater2,Jiang30102007,RevModPhys.83.33}.
To make forward progress, these schemes depend on receiving messages
indicating the success or failure of entangled state creation, and are
known as acknowledged entanglement control~(AEC)~\cite{aparicio2011protocol,VanMeter_2009_2}
or heralded entanglement generation~(HEG)~\cite{muralidharan2016optimal,takeoka2014fundamental}.

This generation achieves high fidelity by entanglement purification, which
requires bidirectional classical communication (a two-way entanglement
purification protocol, or 2-EPP) and  
achieves Bell pairs between non-neighboring nodes by entanglement
swapping between neighboring links~\cite{swapping}. 
To tolerate decoherence, this generation requires entanglement
purification between remote nodes, therefore the cost increases as a
polynomial multiple of  the number of hops. 
(2) The second generation creates encoded Bell pairs between each pair of neighboring nodes
and executes entanglement swapping between encoded Bell pairs to
create encoded Bell pairs between two arbitrary nodes 
~\cite{VanMeter_2009,1367-2630-15-2-023012,knill96:concat-arxiv}.
Encoded Bell pairs over each link are created
by using transversal teleportation-CNOT gates, which consume physical Bell pairs and
require only unidirectional classical communication
(a one-way entanglement purification protocol, 1-EPP).
At the link level this generation is employed only when the photonic qubit's error rate is low, because
the benefit of 1-EPP is ruined if physical level bidirectional entanglement purification (2-EPP) is required.
Fowler et al.'s Bell pair creation for the surface code is also in this generation ~\cite{PhysRevLett.104.180503}.
Their method creates a surface code lattice which spans all the repeater nodes in a path 
between the two end nodes by measuring stabilizers separated in neighboring nodes
by consuming Bell pairs.
Next, the portions of the lattice held by repeater nodes are measured out and 
a Bell pair encoded on the surface code is left between the two end nodes.
%
(3) The third generation directly sends pulses on which a quantum state is encoded by quantum error correction,
which requires unidirectional classical communication (1-EPP) ~\cite{munro2010quantum,munro2012quantum}.
This generation requires very high reception probability to allow direct transmission of states.
We call (1) the entanglement swapping~(ES) model and call (2) and (3) quantum error correction~(QEC) models.

Any generation and model assumes that every node functions properly, but in the real world,
malfunctions and hijacking of nodes are problematic.
A world-wide internetwork will have too many nodes to avoid such problems.
Malfunctions can be detected by self-check of every node, however,
hijackings generally accompany concealment, hence it is difficult to
find them by self-checks alone.

Repeater hijacking is one of the highest priority problems
in network operation, due to its possible impact  on the stability of
the network itself.
In contrast, use of vulnerable applications is a lesser concern from
the point of view of network operations, because it affects the
integrity of 
 the nodes using the quantum applications rather than
   the operational integrity of the network.

In classical networks, there exist many types of attacks and taxonomies
of those have been created.
As an example from the perspective of the attack target, in a denial
of service~(DoS) attack, the attacker infringes on the availability of
information services. An attacker can also listen in on a
communications session. If the listening is passive, the attacker is
referred to as an eavesdropper. An attacker that modifies messages
between the sender and receiver is known as a man in the middle.
In this way, the eavesdropper intercepts messages exchanged
between the two victims and replaces them with other messages, while
letting the victims think that they are directly talking to each other
through a private connection.
A model that has been used in Internet of Things research led to our
model based on the issues of confidentiality (whether data has been
disclosed), integrity (whether data has bee modified), and
availability~\cite{suzuki2015classification}.
Classically, the impact of failures of nodes or attacks on nodes,
especially distributed denial of service attacks, has been
analyzed~\cite{mahadevan2006ilt,mahadevan:_degree_correl,li2004fpa,
  JohnCDoyle10112005}. 
Because attacks that prevent proper use of some nodes can be tolerated by dynamic routing,
the robustness of the network against failures and attacks depends on the structure of the network. 
In earlier work, we proposed a taxonomy of the physical attacks on
individual repeater nodes~\cite{suzuki2015classification}.
For the first step of quantification of those attacks, in this paper we focus on the
impact of attacks on quantum repeaters. 

In this paper, we propose an algorithm to detect hijacked nodes,
similar to quantum key distribution (QKD), utilizing a state
  analysis method such as quantum state tomography~\cite{altepeter2005photonic}.
By consuming a portion of the created Bell pairs for state analysis and
adding some classical communication, we can find abnormal Bell pairs,
whether caused by changes to the noise level, hardware or software
failures, or attacks.
By doing this we can detect man in the middle attacks on the quantum
repeater network, as described in Sec.~\ref{MITM}.
Next we analyze the impact of hijacking on the performance of the overall network.

\begin{figure}[h] 
\centerline{\includegraphics[width=0.65\textwidth]{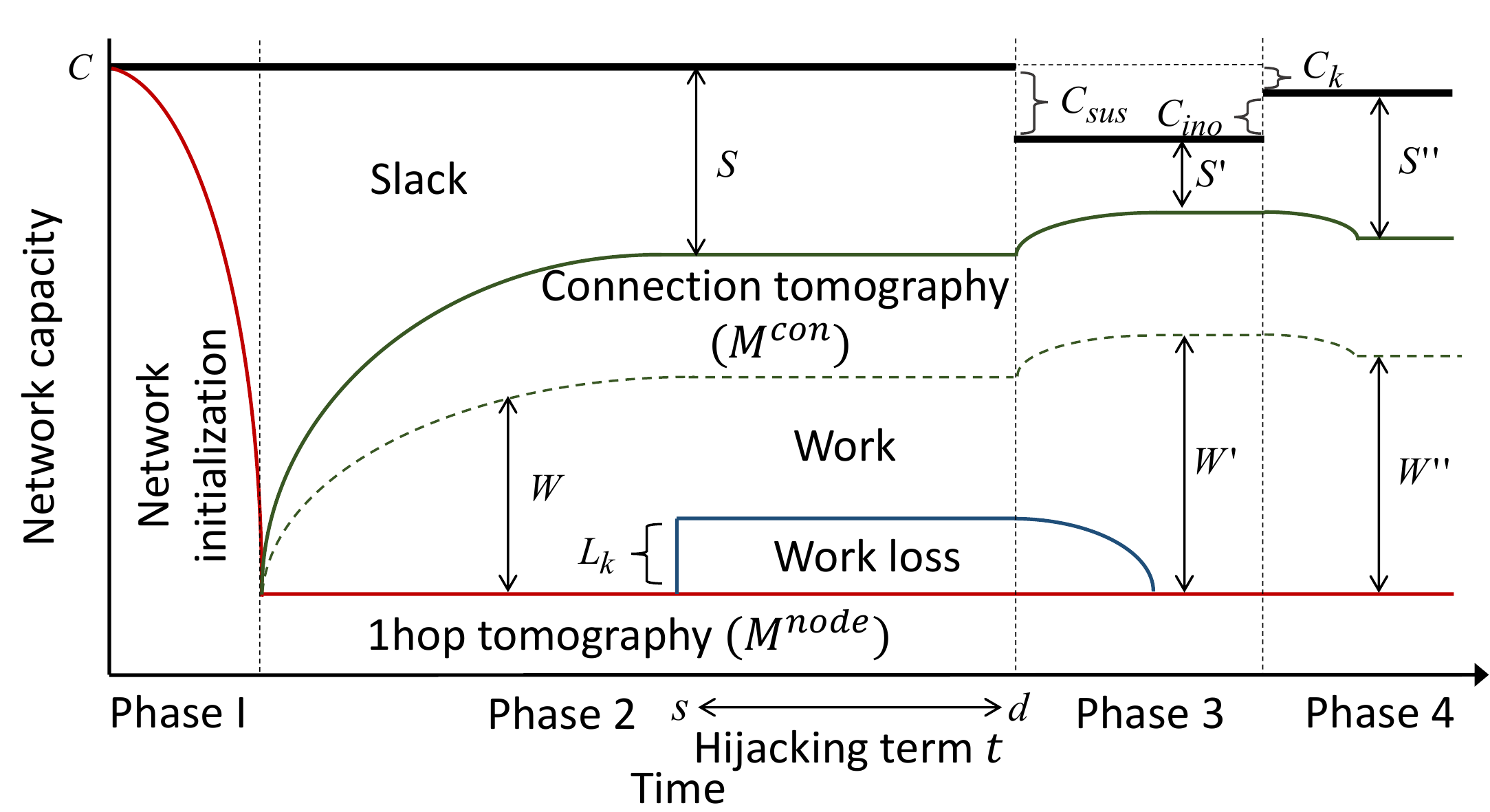}}
\caption{Network slack $S$ over time. We evaluated the
  unavoidable operating costs in each phase (Phase 1. Network
  launching, Phase 2. Normal operation, Phase 3. Response to the
  detection of repeater hijacking, and Phase 4. Back to normal operation.). We show details of this diagram in
  following sections.} \label{fig:MR}
\end{figure}

To discuss the margin of network capacity and the cost for
hijacking detection, we take notice of the
frequency of state analysis for network 
maintenance. The state analysis interval is closely allied to the stability
and integrity of the quantum repeater network.
As the total work approaches the network capacity, queueing delays
grow quickly~\cite{kleinrock1974qsv1}.
Therefore, a network is always designed with a certain amount of ``slack''
capacity, to reduce queueing delays and the probability of connection
failure due to insufficient capacity as the work varies over time.
Fig.~\ref{fig:MR} qualitatively shows the different uses of network
bandwidth, and how they vary over time. These uses will be detailed in Sec.~\ref{Phases_no}
To quantify this discussion, we evaluate the number of entanglement
attempts (quantum optical pulses, e.g. single photons) required
for tomography between nonlocal nearest neighbor repeaters.
We expand this element to include the costs for hijacking detection on ES model and QEC
model repeater networks.

Sec.~\ref{Background} shows the components of a quantum repeater network
and tomography.
In Sec.~\ref{Impact}, we quantify the effects of repeater hijacking.
In Sec.~\ref{Preventing}, we quantify the work of tomography for
hijacking detection and discuss the occurrence of work shedding.
Sec.~\ref{sec_framing} shows how a single hijacker can leverage its position in the
network to dramatically expand an attack and bring down an entire
network by framing other repeaters, leading network administrators to
conclude that an entire set of repeaters has been hijacked and causing
them to isolate enough nodes to partition the network. It then
proposes a solution. 
Sec.~\ref{sec_postprocess} describes the process of network operation
and recovery from hijack detection.  
Sec.~\ref{Discussion} summarizes our results and discusses remaining problems.